\documentclass[aip,apl,reprint,twoside,floatfix,superscriptaddress,nolongbibliography,a4]{revtex4-1} 

\usepackage{graphicx}
\usepackage{amsmath}
\usepackage{amssymb}
\usepackage{amsfonts}
\usepackage{dcolumn}
\usepackage{epsfig}
\usepackage{bm}

\begin{document}

\title{High Q-factor Sapphire Whispering Gallery Mode Microwave Resonator at Single Photon Energies and milli-Kelvin Temperatures}

\author{Daniel L. Creedon}
\email{creedon@physics.uwa.edu.au}
\affiliation{ARC Centre of Excellence for Engineered Quantum Systems, School of Physics, University of Western Australia, 35
Stirling Hwy, Crawley 6009, Western Australia}

\author{Yarema Reshitnyk}
\affiliation{ARC Centre of Excellence for Engineered Quantum Systems, School of Mathematics \& Physics, University of
Queensland, St. Lucia 4072, Queensland, Australia}

\author{Warrick Farr}
\affiliation{ARC Centre of Excellence for Engineered Quantum Systems, School of Physics, University of Western Australia, 35
Stirling Hwy, Crawley 6009, Western Australia}

\author{John M. Martinis}
\affiliation{Department of Physics, University of California, Santa Barbara, California 93106 USA}

\author{Timothy L. Duty}
\altaffiliation{Present affiliation: School of Physics, The University of New South Wales, Sydney 2052, Australia}
\affiliation{ARC Centre of Excellence for Engineered Quantum Systems, School of Mathematics \& Physics, University of
Queensland, St. Lucia 4072, Queensland, Australia}

\author{Michael E. Tobar}
\affiliation{ARC Centre of Excellence for Engineered Quantum Systems, School of Physics, University of Western Australia, 35
Stirling Hwy, Crawley 6009, Western Australia}

\date{\today}

\begin{abstract}
The microwave properties of a crystalline sapphire dielectric whispering gallery mode resonator have been measured at very low excitation strength ($E/\hbar\omega \approx 1$) and low temperatures ($T \approx 30$ mK). The measurements were sensitive enough to observe saturation due to a highly detuned electron spin resonance, which limited the loss tangent of the material to about $2\times10^{-8}$ measured at 13.868 and 13.259 GHz. Small power dependent frequency shifts were also measured which correspond to an added magnetic susceptibility of order 10$^{-9}$. This work shows that quantum limited microwave resonators with $Q$-factors $> 10^8$ are possible with the implementation of a sapphire whispering gallery mode system.
\end{abstract}

\maketitle 

Single crystal sapphire resonators ($\alpha$-Al$_2$O$_3$) are particularly useful in a range of precision microwave experiments due to their extremely low dielectric loss tangent \cite{TobarBlairTransactions1994,cuthbertson,tobarAPBLO,Tobar2000520,Pyb2005apl,PhysRevLett.95.040404,Hartenett2006apl,floch:142907,ivanovtobarTUFFC2009,PhysRevD.82.076001,PhysRevD.81.022003}. Key to the operation of cryogenic devices is the very high electronic $Q$-factors of larger than $10^9$ in whispering gallery (WG) modes at liquid helium temperature, which operate typically at input powers of order 1 mW or above. In fact, the bulk electronic properties of sapphire have been characterized extensively over a wide range of temperatures from room to superfluid liquid helium temperatures using the WG mode technique \cite{tobar:1604,JerzyMTT,JerzyMST}. Recently we extended these tests to temperatures as low as 25 mK, publishing the first observation of electromagnetically induced thermal bistabilty in bulk sapphire due to the material $T^3$ dependence on thermal conductivity and the ultra-low dielectric loss tangent \cite{creedonPRB2010}.
\begin{figure}[b]
\centering
\includegraphics[width=2.49in]{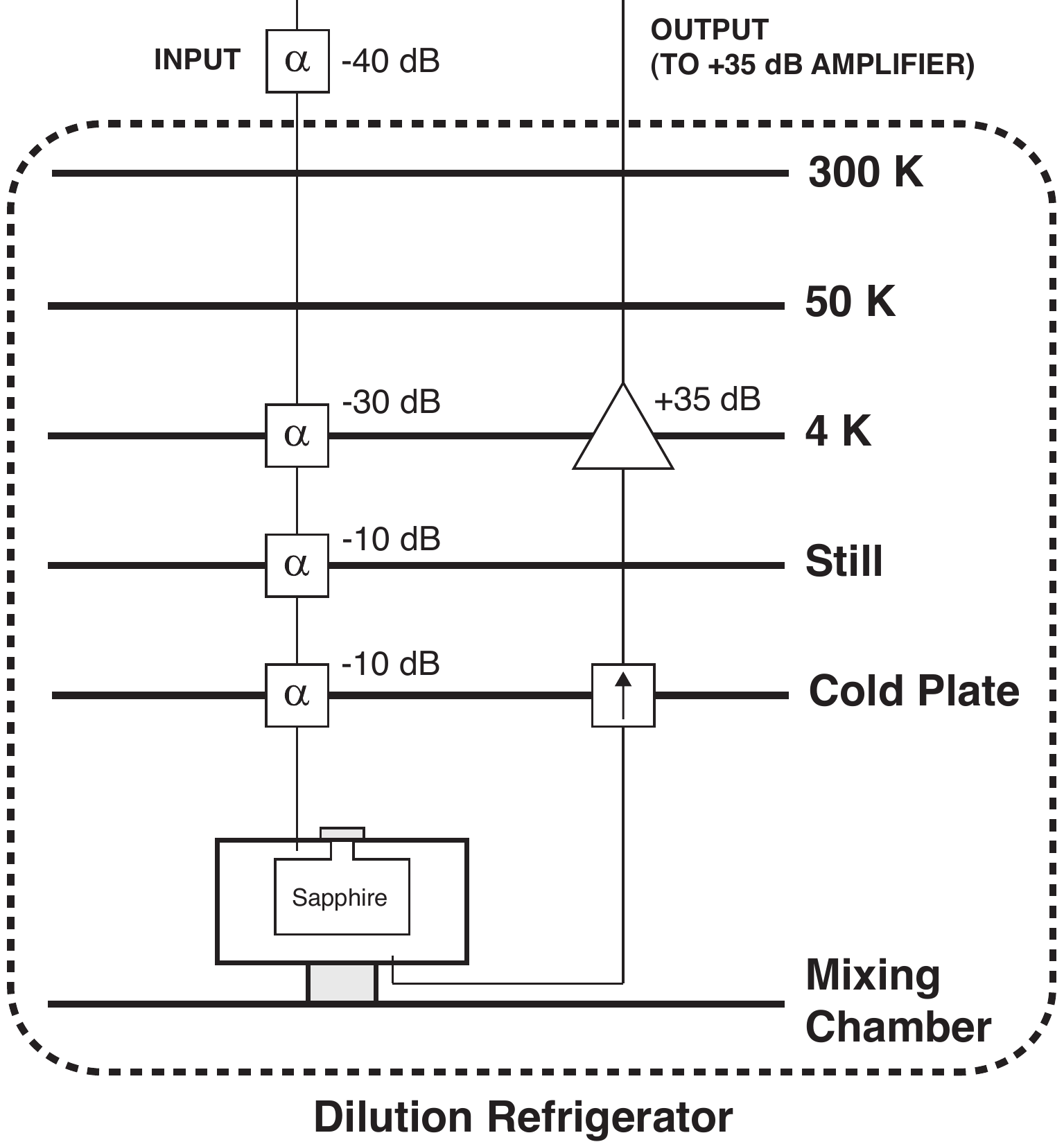}
\caption{\label{figure1}Schematic of the set-up to measure Q-factor and frequency at low temperature and low input power.}
\end{figure}

In recent years, Martinis et al. have shown that dielectric loss from ubiquitous two-level fluctuators, whether they be material defects, microscopic degrees of freedom, or otherwise, is a dominant source of decoherence in superconducting qubits \cite{PhysRevLett.95.210503,o'connell:112903,wang:233508}. For instance, in Josephson junction qubits, dielectric loss in the insulating layer can lead to short coherence times and thus it is very important to improve this parameter \cite{cicak:093502,paik:072505,PhysRevB.80.132501,wang:233508}.  It has been shown that the loss is well modelled by resonant absorption of ``two level states" (TLS).   Longer coherence times can be achieved in part by selecting or engineering insulating materials with superior dielectric loss tangent \cite{barends:023508}. The low loss tangent of monolithic sapphire is well known, but has typically been measured in a high power regime ($P_{inc} = -40$ to +20 dBm) at which TLS are saturated. In this summary, we report on the first measurements of sapphire $Q$-factor at single photon input powers ($P_{inc} \approx -140$ dBm at 13 GHz) to determine the suitability for use in quantum measurement applications.

A highest purity HEMEX-grade sapphire resonator (5cm diameter, 3cm height) was mounted in a silver-plated copper cavity, and affixed to the mixing chamber of a dilution refrigerator and cooled to 25 mK.  A network analyser, locked to a high-stability quartz reference, was used to generate a microwave signal, which was heavily attenuated, injected into the resonator, and amplified back to detectable levels. The Q-factor of the Whispering Gallery modes WGH$_{19,0,0}$ and WGH$_{20,0,0}$ were measured at a range of temperatures and input powers.
\begin{figure*}[t]
\centering
\includegraphics[width=6in]{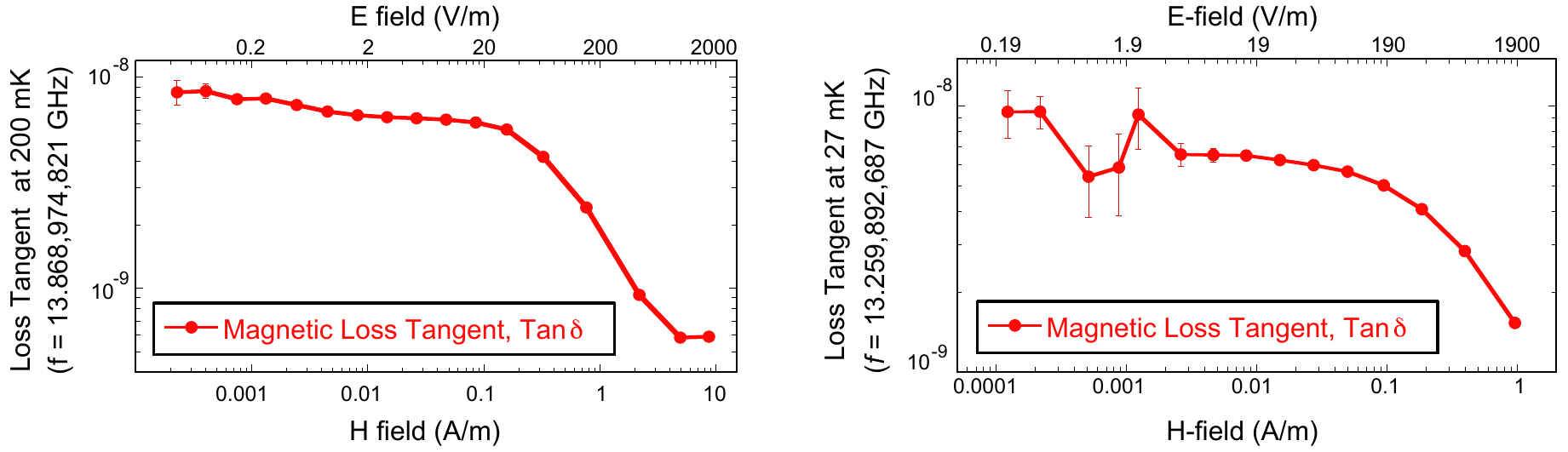}
\caption{\label{fig3}Measured loss tangent versus the $\vec{\mathbf{E}}$ and $\vec{\mathbf{H}}$ field amplitudes for the WGH$_{20,0,0}$ mode (left) and WGH$_{19,0,0}$ mode (right)}
\end{figure*}

As shown in Fig. \ref{figure1}, 50 dB of attenuation was attached to the input line in the cryogenic environment (at or below 4 K), and an additional 40 dB of attenuation was attached at room temperature outside the fridge. The loss in the cables between the Network Analyser and the input probe of the cavity was nominally 7 dB, and the coupling of the probe of 0.02 leads to a further loss of 16.99 dB.  Hence the power at the resonator was adjusted in the analysis to account for an additional 24 dB loss (not marked as attenuation on the schematic in Fig. \ref{figure1}). On the output line, 35 dB of amplification was inserted within the cryogenic environment, a commercial device from JPL, model R3C3M1 CIT-4254-077 (0.5-11 GHz). A second 35 dB amplifier was inserted at room temperature, Low Noise Factory model LNF-LNR1-11B (1-11 GHz).  It should be noted that the resonant modes of interest were in the region of 12-14 GHz, whereas the cryogenic amplifiers were specified for operation below this range. The gain was nominally unaffected, however the noise figure almost doubled at these higher frequencies, severely degrading the signal to noise ratio.

\begin{figure*}[t]
\centering
\includegraphics[width=5.8in]{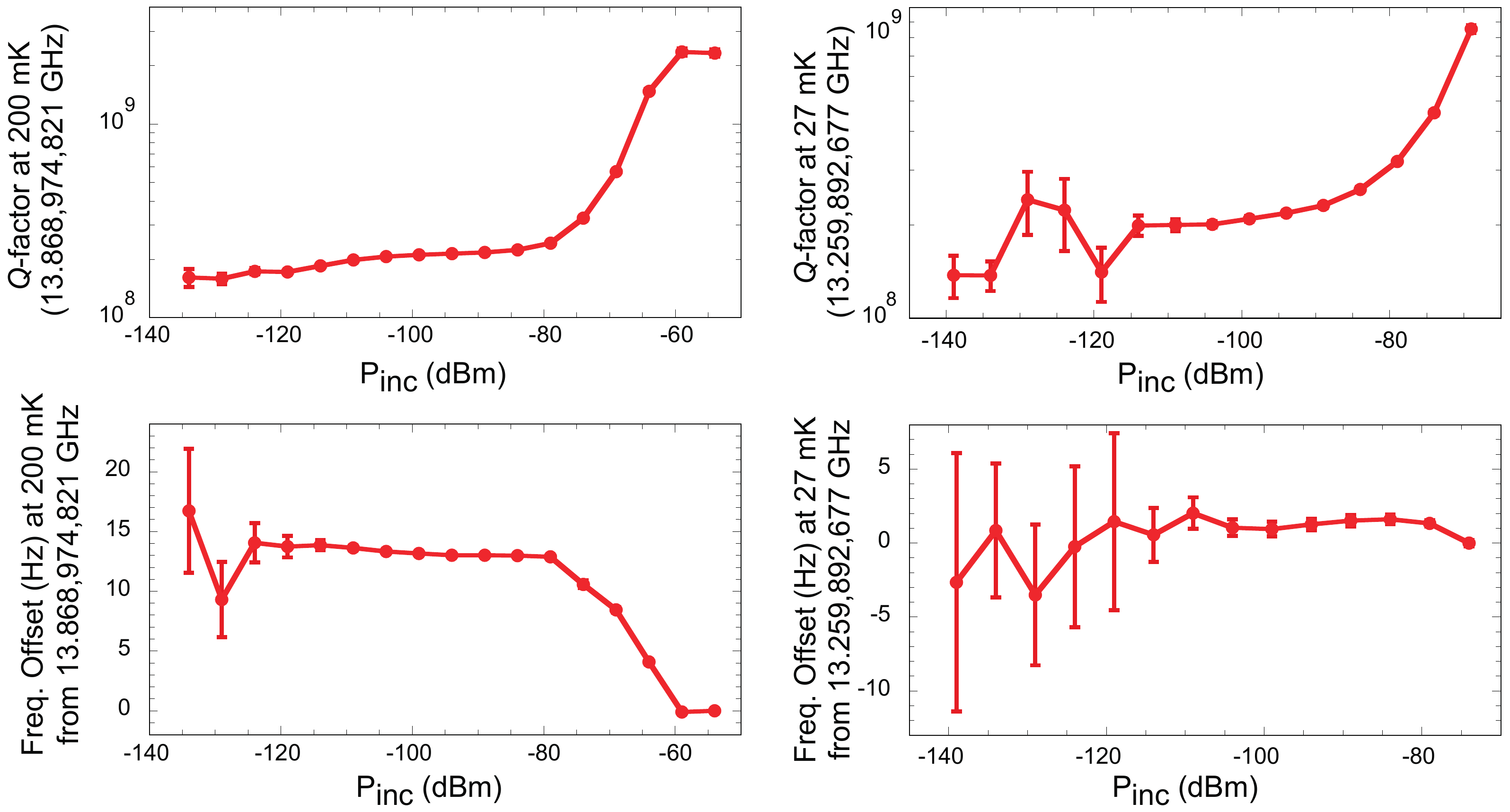}
\caption{\label{figure2}$Q$-factor and frequency shift as a function of incident power for the WGH$_{20,0,0}$ and WGH$_{19,0,0}$ modes at 200 and 27 mK respectively (as labelled)}
\end{figure*}

The Q-factors were determined by fitting a Fano resonance \cite{RevModPhys.82.2257} (Lorentzian profile with an asymmetry factor) to each curve and extracting the linewidth and centre frequency, from which $Q$ can be calculated. Note that at low input powers, signal to noise was degraded and all fits show the standard error (SNR=1) calculated from the Fano resonance fits, which clearly increases at low input powers.
The Q-factor and frequency shift were measured as a function of input power for two modes: $m=19$ ($\sim$13.259 GHz), and $m=20$ ($\sim$13.869 GHz) at 27mK and 200mK respectively, and is shown in Fig. \ref{figure2}. Because the signal-to-noise ratio degrades with incident power, a noticeable increase in the standard error of the $Q$ and frequency determination is observed at powers below -115 dBm.
At higher powers (greater than 1 mW), the biggest effects on mode frequency and $Q$-factor have previously been shown to be due to residual paramagnetic impurities such as Fe$^{3+}$, Cr$^{3+}$ and Ti$^{3+}$ of order parts per billion to part per million \cite{0022-3727-30-19-016,hartnettEL1998,hartnettTUFFC1999,PhysRevD.67.062001,PhysRevB.75.024415,PhysRevLett.100.233901,PhysRevB.79.174432,0957-0233-21-2-025902}. The susceptibility added by these impurities may create frequency-temperature turnover points, which make possible the production of stable frequencies by implementing a high-$Q$ sapphire whispering gallery mode resonator as the frequency-determining device in an active system.  In addition, high-$Q$ cryogenic masers based on concentrations of only parts-per-billion of paramagnetic impurities have been created. For example, when a WG mode is tuned to an electron spin resonance in sapphire such as the Fe$^{3+}$ resonance at 12.04 GHz, large saturation effects are observed and if one pumps at a higher frequency population inversion is created and masing occurs \cite{Pyb2005apl,0957-0233-21-2-025902}.

These results use the same maser sapphire crystal in Bourgeois et al.\cite{Pyb2005apl}, which means that we are most likely interacting with an extremely small number of Fe$^{3+}$ ions in the tail resonance at 12.04 GHz. The susceptibility change upon saturation may be calculated using:
\begin{equation}
\chi^{\prime}=\frac{2}{p_{m_{\bot}}}\frac{\Delta\nu}{\nu}
\end{equation}
Here $p_{m_{\bot}}$ is the magnetic filling factor, and $\Delta\nu/\nu$  the fractional frequency shift (which in general is complex). This gives, for example, susceptibility of order several times 10$^{-9}$ at very low incident power, which begins to saturate at about -90 dBm and achieves full saturation at -60 dBm as shown in Fig. \ref{figure2}.

The reduction in $Q$-factor is thus due to an extra component of magnetic loss (or imaginary susceptibility) supplied by the paramagnetic impurity as a function of input power. Fig. \ref{fig3} shows the calculated electric and magnetic field strength within the resonator as a function of the magnetic loss tangent. This achieved using a separation of variables technique to determine the mode field patterns \cite{106549,PhysRevB.79.174432}, then using standard relations that express the electric and magnetic field energy stored in the resonator to the parameters of the resonator, such as input power $Q$-factor, filling factors and couplings.

As the power decreases, $Q$-factors drop by a factor of between 2 and 10 but remain as high as several hundred million. It is likely that the effect is magnetic, as parametric ions at parts per billion have a dominant effect at low temperatures. Even a long way from the electron spin resonance we see small effects by virtue of the high precision of these measurements. Typically, susceptibility of parts in 10$^{-9}$ added by a detuned ESR is normal, which is saturated at higher power. As the power decreases, the loss tangent tends towards a value of several times 10$^{-9}$. Several factors in the experimental setup limited the quality of results, the most significant being the use of a cryogenic amplifier out of its specified range resulting a reduced signal-to-noise ratio and thus a reduced accuracy of $Q$ determination for the lowest power measurements, hence the large error bars representing the standard error of the fit. Nevertheless, we have shown that high $Q$-factors of parts in 10$^8$ are possible at the energies of a single photon and at milli-Kelvin temperature, which could be useful for a host of quantum measurement applications.

\section*{Acknowledgements}
The authors wish to thank the Australian Research Council for supporting this work under grant numbers FL0992016 \& CE11E0082 for MET, and DP0986932 \& FT100100025 for TLD. JMM is supported by IARPA under ARO award W911NF-09-1-0375.

%

\end{document}